%% file: dual.tex
\documentclass[sigconf,nonacm]{acmart}

\AtBeginDocument{%
  \providecommand\BibTeX{{%
    \normalfont B\kern-0.5em{\scshape i\kern-0.25em b}\kern-0.8em\TeX}}}

\setcopyright{none}
\copyrightyear{}
\acmYear{}
\acmDOI{}

\acmConference[]{}{}{}
\acmBooktitle{}
\acmPrice{}
\acmISBN{}

\acmSubmissionID{514}

\usepackage{subfigure}
\usepackage{multirow}
\usepackage{cleveref}
\usepackage{math}
\input{macro.tex}

\begin{document}

\title{Dual Distribution Alignment Network for \\
Generalizable
Person Re-Identification}


\author{Peixian Chen}
\affiliation{\institution{Xiamen University}}
\email{pxchen@stu.xmu.edu.cn}

\author{Pingyang Dai}
\affiliation{\institution{Xiamen University}}
\email{pydai@xmu.edu.cn}

\author{Jianzhuang Liu}
\affiliation{\institution{Huawei Noah's Ark Lab}}
\email{liu.jianzhuang@huawei.com}

\author{Feng Zheng}
\affiliation{\institution{SUSTech}}
\email{zfeng02@gmail.com}

\author{Qi Tian}
\affiliation{\institution{Huawei Cloud \& AI}}
\email{tian.qi1@huawei.com}

\author{Rongrong Ji}
\affiliation{\institution{Xiamen University, China}}
\email{rrji@xmu.edu.cn}


\input{./1_abstract_dual.tex}




\maketitle

\input{./2_introduction_ji_1.13.tex}
\input{./3_related_eccv}
\input{./4_method.tex}
\input{./5_experiments.tex}
\input{./6_conclusion.tex}

\clearpage
\bibliographystyle{ACM-Reference-Format}
\bibliography{dual}

\end{document}

%% file: macro.tex
\usepackage{xcolor}
\newcommand{\red}[1]{\textcolor{black}{#1}}

\newcommand{\black}[1]{\textcolor{black}{#1}}

\newcommand{\Dataset}{\mathcal{D}}

\newcommand{\Encoder}{E}
\newcommand{\EncoderParameter}{{\theta_e}}
\newcommand{\Transformer}{M}
\newcommand{\TransformerParameter}{{\theta_m}}
\newcommand{\DomainDiscriminator}{D}
\newcommand{\DomainDiscriminatorParameter}{{\theta_d}}
\newcommand{\IdentityDiscriminator}{I}
\newcommand{\IdentityDiscriminatorParameter}{{\theta_i}}
\newcommand{\FeatureExtractor}{F}
\newcommand{\FeatureExtractorParameter}{{\theta_f}}

\newcommand{\LossIDE}{\mathcal{L}_{\mathrm{IDE}}}
\newcommand{\LossTriplet}{\mathcal{L}_{\mathrm{Triplet}}}

\crefdefaultlabelformat{#2\!#1#3}
\crefname{figure}{Fig.}{Figs.}
\Crefname{figure}{Fig.}{Figs.}
\crefname{equation}{Eq.}{Eqs.}
\Crefname{equation}{Eq.}{Eqs.}
\crefname{table}{Tab.}{Tabs.}
\Crefname{table}{Tab.}{Tabs.}
\crefname{section}{Sec.}{Secs.}
\Crefname{section}{Sec.}{Secs.}

\usepackage{pifont}
\newcommand{\cmark}{\ding{51}}%
\newcommand{\xmark}{\ding{55}}%

\usepackage[group-separator={,}]{siunitx}

%% file: 1_abstract_dual.tex
\begin{abstract}
Domain generalization (DG) is promising to handle person Re-Identification (Re-ID), which trains the model using labels from the source domain alone, and then directly adopts the trained model to the target domain without model updating.
However, existing DG approaches are still defected when facing serious domain variations. 
Therefore, DG highly relies on designing domain-invariant features, which is still an open problem, since most existing approaches directly mix multiple datasets to train DG models without considering the inter-domain similarities, \emph{i.e.}, 
examples that  are very similar but from different domains.
%
In this paper,
we present a Dual Distribution Alignment Network (DDAN), which maps 
images into a domain-invariant feature space by selectively aligning the distributions of multiple source domains.
To this end, an alignment network is designed with dual-level constraints, \emph{i.e.}, a novel domain-wise adversarial feature learning and an identity-wise similarity enhancement.
We evaluate our DDAN on a large-scale Domain Generalization Re-ID (DG Re-ID) benchmark. Quantitative results demonstrate that the proposed DDAN can well align the distributions of multiple domains with serious variations, and significantly outperform all existing domain generalization approaches.
\keywords{Person Re-Identification, Domain Generalization}
\end{abstract}

%% file: 2_introduction_ji_1.13.tex
\begin{figure*}[t]
	\centering
	\includegraphics[width=\linewidth]{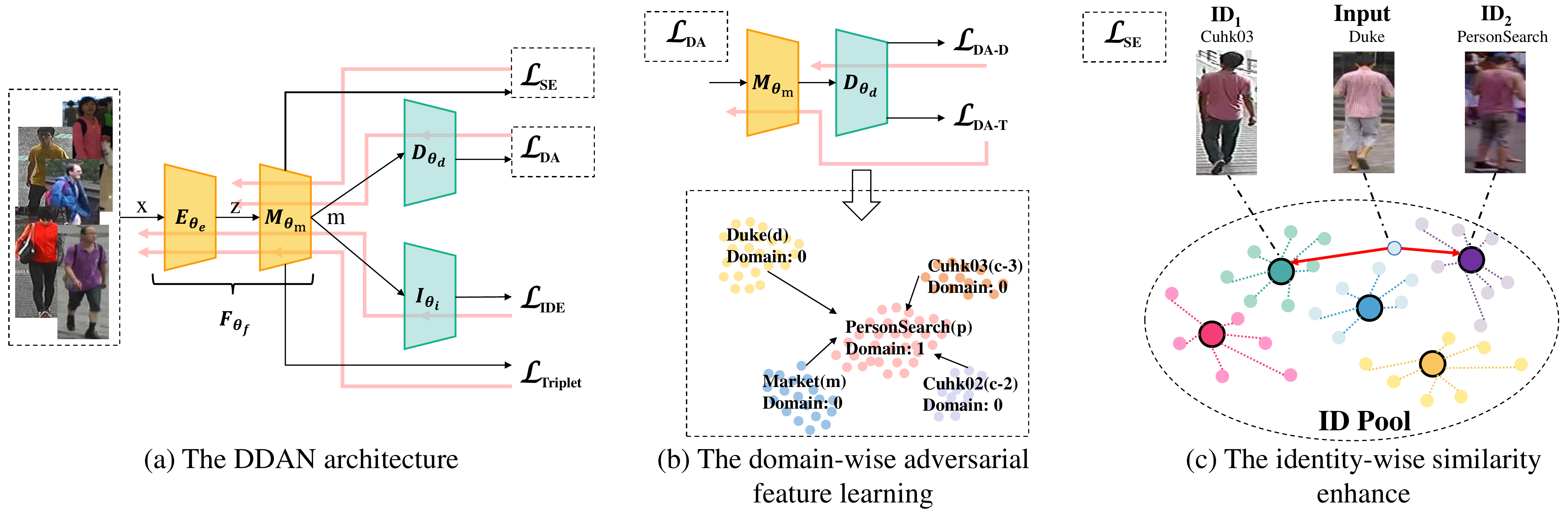}
	\caption{An overview of the proposed DDAN. \textbf{Left:} The network architecture with gradient back-propagation denoted by the red arrows. \textbf{Middle:} Domain-wise adversarial feature learning, where the peripheral domains (labeled as 0) are aligned to the central domain (labeled as 1). \textbf{Right:} Identity-wise similarity enhancement with an ID pool to enhance the inter-domain similarity of feature distributions. Big dots denote the summarized representation of each ID, to which the similarity of new inputs should be selectively enhanced (detailed in \cref{sec:idwise}).}
	\label{fig:network}
\end{figure*}
\section{Introduction}

%
%
Person Re-Identification (Re-ID) aims to identify the same pedestrian captured by different cameras under variant viewpoints, lighting and locations, which has attracted extensive research focus in recent years. 
Along with the success of deep Convolution Neural Networks (CNNs)~\cite{conf/cvpr/Bak017,conf/cvpr/BaiBT17,conf/ijcai/LiZG17}, person Re-ID approaches have achieved remarkable performance when the training and test data are \emph{i.i.d.}~sampled.
Such a setting, however, is indeed problematic in two folds: First, it is prohibitively expensive to collect and label a sufficient amount of training data in the same distribution as the test data, as the testing scenario may not be directly available during model training. Second, the learned deep features naturally  overfit to the training dataset (source domain), which might perform poorly when being directly applied to an unseen domain. 
%
To this end, unsupervised domain adaptation (UDA) has been introduced in person Re-ID. In UDA, a model is learned using data from both the labeled source domain and the unlabeled target domain, which fits the data distribution in the target domain without the cost of labeling~\cite{conf/cvpr/WangZGL18,conf/bmvc/LinLLK18,conf/eccv/BakCL18}. However, as validated in the existing endeavors~\cite{conf/cvpr/PengXWPGHT16,conf/cvpr/Deng0YK0J18,Zhong_2019_CVPR,Fu_2019_ICCV}, UDA approaches require a large amount of target data to achieve satisfactory results.

Comparing to UDA, domain generalization (DG) provides a more preferable real-world setting, which trains a model using multiple source domains and expects this model to perform well in an unseen target domain without any model updating, \emph{i.e.}, adaptation or retraining. 
From this perspective, DG is arguably harder as it does not need any data from the target domain. An optimal DG is supposed to learn a feature representation that is discriminative for the underlying Re-ID task, which should be meanwhile insensitive to the variation of domain distributions.
%
One promising approach is the adversarial feature alignment~\cite{conf/cvpr/LiPWK18,conf/eccv/LiTGLLZT18}, which trains a domain-invariant model by reducing the pair-wise discrepancy between domains with either Maximum Mean Discrepancy (MMD)~\cite{journals/corr/TzengHZSD14} or Jensen-Shannon Divergence (JSD)~\cite{journals/tit/Lin91}.
However, existing DG methods are designed for the classification task.
In contrast, person Re-ID handles the retrieval task, which has fundamental difference\footnote{
DG in classification considers the same set of labels in all source domains. In contrast, DG in person Re-ID/retrieval should compare the feature similarity between  different IDs (corresponding to different labels between source and target domains).}.
To conduct DG in person Re-ID, \cite{conf/cvpr/SongYSXH19} introduced meta-learning to learn a classifier from the gallery images, which outputs matching scores by the dot product between features of the gallery and the probe images. For another instance,
DualNorm~\cite{journals/corr/abs-1905-03422} adopts normalization layers to learn discriminative features that are insensitive to domain variations for similarity matching.

Two fundamental challenges still exist in current DG based person Re-ID, \emph{i.e.}, 
domain-wise variations and identity-wise similarities.
If one directly applies classification based DG methods to person Re-ID, the pair-wise domain alignment might hurt the Re-ID accuracy due to the significant variations between source domains, as there is a statistical trade-off between domain-invariance and classification accuracy~\cite{journals/corr/abs-1904-12543}.
%
On the other hand, the existing DG methods for retrieval (Re-ID) is also problematic, which directly mix all source domains without considering the \emph{inter-domain} similarities\footnote{In this paper, \emph{inter-domain} similarities refer to examples that are similar but from different domains.} between IDs in different domains.
It contradicts most real-world scenarios, as two pedestrian images from different domains with similar visual features are more likely to be incorrectly identified as the same ID.
Therefore, further removal of the domain gap between identities is required.

To address the above two challenges, we propose an end-to-end Dual Distribution Alignment Network (DDAN), which aims to  
reduce the domain-shift among multiple source domains by imposing \emph{dual-level} constraints.
At the domain level, a novel domain-wise adversarial feature learning is proposed to align the feature distributions of different domains. Unlike previous methods~\cite{conf/cvpr/LiPWK18,conf/eccv/LiTGLLZT18} that pair-wisely align all source domains, we selectively reduce the discrepancy between a \emph{central} domain and each of the \emph{peripheral} domains. We formalize these two kinds of domains using the Wasserstein distance~\cite{journals/corr/ArjovskyCB17} (detailed in \cref{sec:domain-wise}), such that
the required distributional shift~\cite{journals/corr/abs-1904-12543} for alignment is minimized.
At the identity level, we enhance the \emph{inter-domain} similarity of features that are similar but from different domains. 
As shown in \cref{fig:network}, the domain-invariance is enforced since we reduce the distance of each example to its top-$k$ similar examples in other domains. 
Beyond the above domain- and identity-level constraints, common constraints in Re-ID, like Identity-Discriminative Embedding (IDE)~\cite{journals/corr/ZhengYH16} and triplet losses, can be also easily integrated into the proposed DDAN model to improve the effectiveness of the learned features. 


\black{To summarize, the proposed Dual Distribution Alignment Network (DDAN) innovates in the following three aspects:
\begin{enumerate}
    \item We propose a novel domain-wise adversarial feature learning scheme. Unlike previous methods, our method aligns domains with minimal distributional shift to mitigate the loss of accuracy. 
    \item We introduce an identity-wise similarity enhancement, where features of identities from different domains should be closer than those from the same domain but with less appearance similarities.
    \item 
    We evaluate our method in a large-scale DG Re-ID benchmark~\cite{conf/cvpr/SongYSXH19}, with comparisons to a variety of alternative and cutting-edge DG approaches~\cite{conf/cvpr/SongYSXH19,journals/corr/abs-1905-03422}. Quantitative results show that DDAN achieves state-of-the-art performance in a large-scale DG Re-ID benchmark, where the rank-1 accuracies on VIPeR, PRID, GRID, and i-LIDS are 56.5\%, 62.9\%, 50.6\%, and 78.5\%, respectively.
\end{enumerate}}

%% file: 3_related_eccv.tex
\section{Related Work}
\textbf{Person Re-identification.}
\red{Existing methods of supervised person Re-ID typically learn a distance metric~\cite{conf/cvpr/KostingerHWRB12,conf/eccv/XiongGCS14,journals/pami/ZhengGX13}, a subspace~\cite{conf/eccv/WangGZX16,journals/pami/ChenZZL18}, or view-invariant discriminative features~\cite{conf/iccv/ZhengSTWWT15,conf/cvpr/LiaoHZL15,conf/eccv/GrayT08}.}
Along with the success of deep CNNs, person Re-ID has achieved remarkable performance under the \emph{i.i.d.}~assumption between training and test data~\cite{conf/cvpr/ChengGZWZ16,conf/cvpr/Paisitkriangkrai15,conf/cvpr/MatsukawaOSS16,conf/aaai/ChenCZH17,conf/cvpr/Bak017,conf/cvpr/BaiBT17,conf/ijcai/LiZG17}. 
However, the learned models commonly overfit to the training dataset, and would perform poorly when being directly applied to unseen datasets.
Therefore, unsupervised domain adaptation (UDA) approaches are proposed for person Re-ID~\cite{conf/cvpr/PengXWPGHT16,conf/bmvc/LinLLK18,conf/cvpr/WangZGL18,conf/eccv/BakCL18}. 
In such a setting, a labeled source domain is involved to help the model to fit the distribution of an unlabeled target domain. 
\red{For instance, Peng \emph{et al.}~\cite{conf/cvpr/PengXWPGHT16} proposed an asymmetric multi-task dictionary learning to learn discriminative representation for the target domain.
Another group of UDA-based person Re-ID methods exploits generative adversarial networks: Deng \emph{et al.}~\cite{conf/cvpr/Deng0YK0J18} employed CycleGAN~\cite{conf/iccv/ZhuPIE17} to translate images from the source domain to the target domain. Zhong \emph{et al.}~\cite{conf/eccv/ZhongZLY18} generated images with different camera styles in the target domain to enforce camera invariance.
Domain alignment is also exploited in recent endeavors. For instance in~\cite{conf/cvpr/WangZGL18}, a transferable model was proposed to jointly learn attribute-identity discriminative representation for the target domain.
Recent works in UDA for person Re-ID also investigated the clustering of data samples in the target domain for similarity measurement~\cite{Zhong_2019_CVPR}, which was later extended to finer-grained with person's part-level features~\cite{Fu_2019_ICCV}.}
However, all the above UDA approaches typically require a large amount of target data (images and attributes) to achieve satisfactory results by avoiding overfitting to the source data.
%

\noindent\textbf{Domain Generalization.}
DG methods aim to learn a generalizable model, which tries to remove the domain-shift without needing the data of the target domain during training. 
In this regard, many methods have been proposed to conduct DG for the classification task.
\cite{conf/eccv/XuLNX14} trains a model for each source domain, and selects the best one for each target domain during the test phase. 
There exist more efficient choices, such as learning a model to extract task-specific and domain-invariant features. For instance, \cite{conf/icml/MuandetBS13} proposed to learn features via kennel-based optimization, \cite{conf/iccv/GhifaryKZB15} learns a multi-task auto-encoder to extract domain-invariant features, \cite{conf/ijcai/YangG13} uses canonical correlation analysis (CCA) as a domain-distance regularization for DG.
Besides, model-agnostic meta-learning~\cite{conf/icml/FinnAL17} was also introduced in~\cite{conf/icml/LiYZH19,journals/corr/abs-1910-13580}.
%
Another group of works adopted adversarial feature alignment~\cite{conf/cvpr/LiPWK18,conf/eccv/LiTGLLZT18} to train a domain-invariant model by reducing the pairwise domain discrepancy with either MMD or JSD.
In response to the popularization of domain-invariance approaches, \cite{journals/corr/abs-1904-12543} revealed that there is a statistical trade-off between domain-invariance and classification accuracy.
As such, we have observed the same problem in person Re-ID: The pairwise domain alignment can reduce the feature discriminability, which however needs extensive distributional shifts to align different Re-ID datasets.

%
To handle DG in person Re-ID, DIMN~\cite{conf/cvpr/SongYSXH19} adopted meta-learning to learn the classifier from the gallery images, which outputs matching scores by the dot product between features of the gallery and the probe images. However, such a meta-learning scheme can increase the complexity of optimization, and will greatly decrease the test speed correspondingly. 
As a more straightforward solution, DualNorm~\cite{journals/corr/abs-1905-03422} employs both 
Instance Normalization (IN)~\cite{journals/corr/UlyanovVL16} 
and Batch Normalization (BN)~\cite{conf/icml/IoffeS15} 
to improve the DG performance of the feature extractor.

Our DDAN is distinguished from the above methods in two folds: At the domain level, instead of adopting the pairwise alignment, we selectively reduce the discrepancy between a \emph{central} domain and the \emph{peripheral} domains, such that the loss of distinguishability is minimized. At the identity level, differing from existing methods that directly mix all source domain data, we also consider the inter-domain similarities between identities in different domains, which is quantitatively proven to be very effective in promoting the feature discriminability.

%% file: 4_method.tex
\section{The Proposed Method}
\textbf{DG for Person Re-ID.}
In the training phase, we access to $M$ datasets (source domains) $\Dataset_1, \Dataset_2, ..., \Dataset_M$. Each dataset has a set of non-overlapping labels as we assume the IDs among different datasets are non-overlapping. In the test phase, the trained model is ``freezed'' and is directly applied to a new unseen dataset (target domain) without further model updating or retraining.
We denote inputs from source domain $s$ as $X_s=\s{(\bm{x}^s_i, y^s_i)}^{N_s}_{i=1}$, where $N_s$ is the number of labeled data in domain $s$. As shown in \cref{fig:network}, the encoder $\Encoder(\bm{x};\EncoderParameter)$ parameterized by $\EncoderParameter$ maps an image $\bm{x}$ to a feature map $\bm{z}$. $\Transformer(\bm{z};\TransformerParameter)$ is a mapping network parameterized by $\TransformerParameter$, which maps the feature map $\bm{z}$ from different distributions to the one in a shared feature space (denoted by $\bm{m}$). For simplicity, we denote the feature extractor as the composition of the encoder and the mapping network, that is, $\FeatureExtractor=\Transformer\circ \Encoder$, which is parameterized by $\theta_f=(\theta_m, \theta_e)$. Then a domain discriminator $\DomainDiscriminator(\bm{m}; \DomainDiscriminatorParameter)$ parameterized by $\DomainDiscriminatorParameter$ is used to distinguish the domain to which the inputs belong to, and an identity discriminator $\IdentityDiscriminator(\bm{m}; \IdentityDiscriminatorParameter)$ parameterized by $\IdentityDiscriminatorParameter$ is used to increase the effectiveness of the learned features. 
Note that in this section, we only write the parameters that are updated through back-propagation on the left-hand side of the following equations.

\subsection{Baseline Configuration}
We consider the common aggregation (AGG) method for the DG problem as our baseline, in which a model is trained using all source domains. In person Re-ID, given the labeled training images, an effective strategy is to learn the IDE,
which often casts the training process to a classification problem using the cross-entropy loss $\ell_\mathrm{CE}$, as shown below:
\begin{equation}
\LossIDE(X; \FeatureExtractorParameter, \IdentityDiscriminatorParameter)=
\frac{1}{n_{bs}}\sum_{n=1}^{n_{b s}}
\ell_\mathrm{CE}\p[\Big]{
  \IdentityDiscriminator\p[\big]{
    \FeatureExtractor\p{
      \bm{x}_n; \FeatureExtractorParameter
    }; \IdentityDiscriminatorParameter
  }, y_n
},
\label{eqn:celoss}
\end{equation}
where $n_{bs}$ denotes the number of samples in a mini-batch $X=\s{(\bm{x}_n, y_n)}_{n=1}^{n_{bs}}$. After training, the feature extractor $\FeatureExtractor_\FeatureExtractorParameter$ is used to extract features from input images. 

As another common and effective criterion for similarity learning in person Re-ID, the triplet loss is often used to shorten the intra-class distance and to widen the inter-class distance~\cite{journals/corr/HermansBL17}. It can be formulated as: 
\begin{equation}
\LossTriplet(X; \FeatureExtractorParameter) =
\sum_{\bm{x}_a\in \FeatureExtractor\p{
      X; \FeatureExtractorParameter
    }}\p[\big]{
  d(\bm{x}_a, \bm{x}_p) - d(\bm{x}_a, \bm{x}_n) + m
},
\label{eqn:tripletloss}
\end{equation}
where $d$ denotes the Euclidean distance, $m$ is the margin, $\bm{x}_a$ denotes the anchor point, and $\bm{x}_p$ and $\bm{x}_n$ are the hardest positive and negative examples corresponding to $\bm{x}_a$, respectively.
In other words, $\bm{x}_p$ is the farthest sample with the same label as $\bm{x}_a$, and $\bm{x}_n$ is the nearest sample with a different label as $\bm{x}_a$ (all within a mini-batch).

\subsection{Domain-wise Adversarial Feature Learning}\label{sec:domain-wise} 
We reduce the overall discrepancy of all source domains to help the model learn a mapping from input images to a domain-invariant feature space. 
%
In this subsection, we adopt adversarial learning to encourage the distributions of various domains to be close to a uniform one. 
\black{Towards an important early exploration, most methods align each pair of source domains to reduce the overall discrepancy, such as~\cite{conf/cvpr/LiPWK18,conf/eccv/LiTGLLZT18}.}
%
However, as validated in \cite{journals/corr/abs-1904-12543}, there exists a statistical trade-off between domain-invariance and classification accuracy. Thus, the pairwise alignment of source domains can have negative impact on the learned features, as these features have to be heavily shifted to align outlying domains.
\red{To explain, some domain in person Re-ID may have an exceptionally different distribution than the other domains. In such a case, the pair-wise alignment will introduce unnecessary distributional shift towards the outlying domain, and finally leads to the decreasing of accuracy.}

To overcome this problem, we choose the most ``generalizable'' domain as the \emph{central} domain and refer the remaining as the \emph{peripheral} domains. \black{Here, the most ``generalizable'' domain should have the distribution that is similar to most of the remaining domains.} In other words, a central domain is the one that minimizes the distributional shift
needed for aligning the other peripheral domains to it. Then, an efficient way of aligning the source domains with the minimum negative impact can be
determined: Instead of pair-wisely aligning every two domains, we only align the peripheral domains to the central one. Such a setting brings a better generalization while avoiding the negative impact as mentioned above. 

We explain our definition of the central and peripheral domains as follows:
%
We employ the Wasserstein distance $d_\mathrm{WS}$\black{, which is used to measure the distance between two distributions,} to quantify the needed distributional shift for aligning two domains. We then define the central domain $c^*$ as:
\begin{equation}
c^*
=\mathop{\arg\min}_{c\in\mathcal{S}} \sum_{i\in\mathcal{S}\setminus{\s{c}}} d_\mathrm{WS}\p[\big]{X_c, X_i},
\label{eqn:wass}
\end{equation}
where 
$\mathcal{S}$ is the set of all source domains,
$X_c$ and $X_i$ are the samples from domains $c$ and $i$, respectively.
%
In this way, all domains are aligned conveniently with the overall distributional shift being minimized. 
Additionally, we define the ``domain label'' of each domain by checking if it is central (1) or peripheral (0). The chosen central domain will be specified in \cref{sec:damain-wise}.

After the determination of the central and peripheral domains, we further describe the adversarial feature learning component,
%
which involves a pair of generator and discriminator. We regard the mapping network $\Transformer_\TransformerParameter$ as the generator, which maps the distributions of the peripheral domains to a uniform one similar to the central domain. 
The discriminator $\DomainDiscriminator_{\DomainDiscriminatorParameter}$ is optimized by minimizing the following cross-entropy loss \black{to correctly distinguish whether an example belongs to the central domain or one of the peripheral domains}:
\begin{equation}
\mathcal{L}_\mathrm{DA-D}(X; \DomainDiscriminatorParameter)
= \frac{1}{n_{bs}} \sum_{n=1}^{n_{bs}}
\ell_\mathrm{CE}\p[\Big]{
  \DomainDiscriminator\p[\big]{
    \FeatureExtractor(\bm{x}_n; \FeatureExtractorParameter); \DomainDiscriminatorParameter
  }, c_n 
},
\label{eqn:discriminator}
\end{equation}
where $X=\s{(\bm{x}_n, c_n)}_{n=1}^{n_{bs}}$ is the input mini-batch with domain labels.
The mapping network $\Transformer_\TransformerParameter$ is trained to fool the discriminator $\DomainDiscriminator_{\DomainDiscriminatorParameter}$ by generating domain-invariant features. It can be achieved by minimizing the negative entropy of the predicted domain distributions with respect to $\TransformerParameter$ as:
\begin{equation}
\begin{aligned}
\mathcal{L}_\mathrm{DA-T}(X; \TransformerParameter) =
-\frac{1}{n_{bs}} \sum_{n=1}^{n_{bs}} \log \DomainDiscriminator\p[\Big]{\FeatureExtractor\p[\big]{\bm{x}_{n}; \FeatureExtractorParameter}; \DomainDiscriminatorParameter}.
\end{aligned}
\label{eqn:transfer}
\end{equation}
The above training process adversarially aligns the feature distributions of the peripheral domains to a uniform one similar to the central domain. \red{Note that, unlike previous methods that pair-wisely align multiple domains, our alignment is conducted in a specific way from the peripheral domains to the central domain. Thus, our method can minimize the negative impact on the learned features when aligning outlying person Re-ID domains.}

\subsection{Identity-wise Similarity Enhancement}
\label{sec:idwise}
The above domain-wise distribution alignment successfully aggregates as many datasets as possible, but does not consider the local relationship between cross-domain instances. To this end, we derive the identity-wise similarity constraint from the real-world scenario: Two pedestrian images with more similar visual features are more likely to be identified as the same person (ID). In other words, the feature embeddings of these two images should be closer than those with less similar visual features, even if these two images are from different domains.



To this end, we accumulate the learned knowledge of the Re-ID model by defining an ID pool, in which we store the representations of all IDs and enhance the distributional similarity between the newly incoming examples and the visually similar IDs in other domains.
We summarize the representation of each ID $\mu$ by computing its running mean representation $\bar{\bm{r}}_\mu$ in an iterative fashion as:
%
\begin{equation}
\bar{\bm{r}}^{(e, t+1)}_\mu = \frac{1}{t+1}
\p[\Big]{t \cdot \bar{\bm{r}}^{(e, t)}_\mu + \FeatureExtractor(\bm{x}; \FeatureExtractorParameter)},
\label{eqn:mean-iter}
\end{equation}
where $\bm{x}$ is an input image of ID $\mu$, and 
the superscript $(e, t)$ denotes the $t$-th update of ID $\mu$ in the $e$-th epoch.
We further accumulate this mean representation over epochs, leading to the final effective representation $\hat{\bm{r}}_\mu$ as:
\begin{equation}
\hat{\bm{r}}^{(e+1, t)}_\mu =
\alpha \cdot \hat{\bm{r}}^{(e, \backslash)}_\mu
+ (1 - \alpha) \cdot \bar{\bm{r}}^{(e+1, t)}_\mu,
\label{eqn:mean-epoch}
\end{equation}
where $\hat{\bm{r}}^{(e, \backslash)}_\mu$ denotes the final mean representation obtained in epoch $e$, and the hyper-parameter $\alpha \in [0,1]$ controls the updating rate. All the variables in \cref{eqn:mean-iter,eqn:mean-epoch} are initialized to zero. 
%
%

The obtained mean representation $\hat{\bm{r}}_\mu$ conveys how samples of a particular ID are generally represented. 
As such, we want to establish the relationship between the representations of each incoming instance and its similar IDs (from different domains) in the ID pool.
%
%
%
%
Since these paired representations are essentially from different domains that have exceptionally unmatched entries due to the domain variations, we cannot directly make them close in some distance metrics (\emph{e.g.}, $\ell_1$ and $\ell_2$). Instead, we use softmax to normalize these features and minimize their symmetric KL-divergence.

Specifically, given an image $\bm{x}_n$ from the peripheral domains, we search for its top-$k$ similar IDs in other domains with the Cosine similarity. For $\bm{x}_n$ from the central domain, we instead search in the same domain to stabilize the distribution of the central domain.  
\red{To explain,  we align the domains in a specific way from the peripheral domains to the central domain, instead of pair-wisely or the other way around.}
Then, we minimize
\begin{equation}
\begin{aligned}
\mathcal{L}_\mathrm{SE}(X; \TransformerParameter) = \sum_{n=1}^{n_{bs}}\br[\bigg]{
\frac{1}{k}\sum^k_{i=1}
\br[\Big]{
\ell_\mathrm{KL}\p[\Big]{\mathrm{sm}\p[\big]{\FeatureExtractor\p{\bm{x}_n; \FeatureExtractorParameter}} \Big\| \mathrm{sm}\p[\big]{\hat{\bm{r}}_i}}
+ \\
\ell_\mathrm{KL}\p[\Big]{\mathrm{sm}\p[\big]{\hat{\bm{r}}_i} \Big\| \mathrm{sm}\p[\big]{\FeatureExtractor\p{\bm{x}_n; \FeatureExtractorParameter}} }
}},
\end{aligned}
\label{eqn:kl}
\end{equation}
where $\ell_\mathrm{KL}(\bm{p}\|\bm{q})=\sum_r p_r \log\p{{p_r}/{q_r}}$ is the KL-divergence, and $\mathrm{sm}(\bm{x})=\mathrm{softmax}(\bm{x}/\tau)$ is the softmax function at temperature $\tau$. 
This setting further eliminates domain-shift identity-wisely and helps to learn domain-invariant features.
%

%
\subsection{The Overall Objective Function}
Recall that DDAN consists of a novel domain-wise adversarial feature learning and an identity-wise similarity enhancement across different source domains. The overall loss function in a training mini-batch $X$ is thus defined as the sum of: 
\begin{equation}
\begin{aligned}
\mathcal{L}_1(X; \FeatureExtractorParameter, \IdentityDiscriminatorParameter) &= \mathcal{L}_\mathrm{IDE}(X; \FeatureExtractorParameter, \IdentityDiscriminatorParameter) + \lambda_1 \cdot \mathcal{L}_\mathrm{Triplet}(X; \FeatureExtractorParameter), \\
\mathcal{L}_2(X; \TransformerParameter) &= \lambda_2 \cdot \mathcal{L}_\mathrm{DA-T}(X; \TransformerParameter) + \lambda_3 \cdot \mathcal{L}_\mathrm{SE}(X; \TransformerParameter), \\
\mathcal{L}_3(X; \DomainDiscriminatorParameter) &= \mathcal{L}_\mathrm{DA-D}(X; \DomainDiscriminatorParameter), 
\end{aligned}
\label{eqn:total}
\end{equation}
where $\lambda_1, \lambda_2$ and $\lambda_3$ are the trade-off parameters. 
For simplicity, we only write parameters that will be updated through back-propagation on both sides of \cref{eqn:total}.
To summarize, we set the above loss functions to learn to embed the input images into a domain-invariant feature space, in which our model can generalize better to new unseen domains. 
%

%% file: 5_experiments.tex

\begin{table*}[tb]
	\centering
	\resizebox{\textwidth}{!}{
		\begin{tabular}{c|c|cccc|cccc|cccc|cccc}
			\hline
			\multirow{2}{*}{Method} & \multirow{2}{*}{Type} & \multicolumn{4}{c|}{VIPeR} & \multicolumn{4}{c|}{PRID} & \multicolumn{4}{c|}{GRID} & \multicolumn{4}{c}{i-LIDS} \\ \cline{3-18} 
			&& R-1 & R-5 & R-10 & \emph{m}AP & R-1 & R-5 & R-10 & \emph{m}AP & R-1 & R-5 & R-10 & \emph{m}AP & R-1 & R-5 & R-10 & \emph{m}AP \\ \hline
			Ensembles~\cite{conf/cvpr/Paisitkriangkrai15} & S & 45.9 & 77.5 & 88.9 & - & 17.9 & 40.0 & 50.0 & - & - & - & - & - & 50.3 & 72.0 & 82.5 & -  \\
			ImpTrpLoss~\cite{conf/cvpr/ChengGZWZ16} & S & 42.3 & 71.5 & 82.9 & - & 29.8 & 52.9 & 66.0 & - & - & - & - & - & - & - & - & - \\
			GOG~\cite{conf/cvpr/MatsukawaOSS16}       & S & 49.7 & \textbf{79.7} & 88.7 & - & - & - & - & - & 24.7 & 47.0 & 58.4 & - & - & - & - & - \\ 
			MTDnet~\cite{conf/aaai/ChenCZH17}    & S & 47.5 & 73.1 & 82.6 & - & 32.0 & 51.0 & 62.0 & - & - & - & - & - & 58.4 & 80.4 & 87.3 & - \\
			OneShot~\cite{conf/cvpr/Bak017}   & S & 34.3 & - & - & -       & 41.4 & - & - & -   & - & - & - & -     & 51.2 & - & - & -  \\
			SSM~\cite{conf/cvpr/BaiBT17}       & S & 53.7 & - & \textbf{91.5} & - & - & - & - & - & 27.2 & - & 61.2 & - & - & - & - & - \\
			JLML~\cite{conf/ijcai/LiZG17}      & S & 50.2 & 74.2 & 84.3 & - & - & - & - & - & 37.5 & 61.4 & 69.4 & - & - & - & - & -  \\
			\hline \hline
			TJAIDL~\cite{conf/cvpr/WangZGL18}    & UDA & 38.5 & - & - & - & 34.8 & - & - & - & - & - & - & - & - & - & - & - \\
			MMFAN~\cite{conf/bmvc/LinLLK18}    & UDA & 39.1 & - & - & - & 35.1 & - & - & - & - & - & - & - & - & - & - & - \\
			Synthesis~\cite{conf/eccv/BakCL18} & UDA & 43.0 & - & - & - & 43.0 & - & - & - & - & - & - & - & 56.5 & - & - & - \\
			\hline \hline
 			DIMN~\cite{conf/cvpr/SongYSXH19}      & DG & 51.2 & 70.2 & 76.0 & 60.1 & 39.2 & 67.0 & 76.7 & 52.0 & 29.3 & 53.3 & 65.8 & 41.1 & 70.2 &\textbf{89.7} & \textbf{94.5} & 78.4 \\
			DualNorm~\cite{journals/corr/abs-1905-03422}  & DG & 53.9  & 62.5 & 75.3 & 58.0 & 60.4  & 73.6 & 84.8 & 64.9 & 41.4  & 47.4 & 64.7 & 45.7 & 74.8  & 82.0 & 91.5 & 78.5 \\ 
			DDAN (Ours)       & DG & 52.3 & 60.6 & 71.8 & 56.4 & 54.5  & 62.7 & 74.9 &58.9 & \textbf{50.6} & \textbf{62.1} & \textbf{73.8} & \textbf{55.7} & \textbf{78.5} & 85.3 & 92.5 & \textbf{81.5}  \\
			DDAN+DualNorm (Ours) & DG &\textbf{56.5} &65.6 &76.3 &\textbf{60.8} &\textbf{62.9} &\textbf{74.2} &\textbf{85.3} &\textbf{67.5} & 46.2 &55.4 &68.0 &50.9 &78.0 &85.7 &93.2 &81.2\\\hline
		\end{tabular}
	}
	\caption{Comparison (\%) against the baselines, where ``R'': rank; ``S'': supervised training with the target dataset; ``-'': no report.}
	\label{table:result}
\end{table*}
\section{Experiments}
\subsection{Datasets and Settings}
\textbf{Datasets.}
We conduct experiments on the large-scale DG Re-ID benchmark in~\cite{conf/cvpr/SongYSXH19} to evaluate our DG model for person Re-ID. Specifically, CUHK02~\cite{conf/cvpr/LiW13}, CUHK03~\cite{conf/cvpr/LiZXW14}, Market-1501~\cite{conf/iccv/ZhengSTWWT15}, DukeMTMC-ReID~\cite{conf/iccv/ZhengZY17} and CUHK-SYSU PersonSearch~\cite{journals/corr/XiaoLWLW16} are taken as the source datasets. All the images in these source datasets, regardless of their train/test splits, are used for training, in total \num{121765} images of \num{18530} identities. The datasets VIPeR~\cite{conf/eccv/GrayT08}, PRID~\cite{conf/scia/HirzerBRB11}, GRID~\cite{journals/ijcv/LoyXG10}, and i-LIDS~\cite{conf/bmvc/ZhengGX09} are used as the target datasets for testing, in which we follow the single-shot setting with the numbers of probe/galley images set to:  $316/316$ on VIPeR, $100/649$ on PRID, $125/900$ on GRID,  and $60/60$ on i-LIDS. 
%

\noindent
\textbf{Settings.}
We implement our model with PyTorch and train it on a single 1080 Ti GPU. The MobileNetV2~\cite{conf/cvpr/SandlerHZZC18} with a width multiplier of $1.0$ is used as the backbone network for the encoder $\Encoder_{\EncoderParameter}$ and mapping network $M_{\theta_m}$. The weights are pretrained on ImageNet. Note that the mapping network $\Transformer_{\TransformerParameter}$ is actually the last convolution layer of MobileNetV2.
The learning rate is initially set to $0.1$ and multiplied by $0.1$ per $40$ epochs. Our domain discriminator $\DomainDiscriminator_{\DomainDiscriminatorParameter}$ consists of a $128$-D and a $2$-D fully-connected (FC) layers with a batch normalization (BN), while the identity discriminator $\IdentityDiscriminator_{\IdentityDiscriminatorParameter}$ is a $\num{18530}$-D (\emph{i.e.}, the number of identities) FC layer with a BN. The updating rate $\alpha$ in \cref{eqn:mean-epoch} is set to $0.05$. The triplet loss margin in \cref{eqn:tripletloss} is \num{0.3}. The $\tau$ of softmax in \cref{eqn:kl} is $2\times10^{-3}$. The weights of the losses in~\cref{eqn:total} are set to $\lambda_1=1.0$, $\lambda_2=0.18$ and $\lambda_3=0.05$. The model is trained for $100$ epochs with a batch size of $64$ (each identity comes with $4$ images). The loss $\mathcal{L}_\mathrm{SE}$ is only enabled after the 4th epoch.
The test results are averaged over $10$ random probe/gallery splits.

\subsection{Comparison to the State-of-the-Arts} 
As shown in~\cref{table:result}, we compare DDAN with other methods on VIPeR, PRID, GRID and i-LIDS. The compared methods include 7 supervised training (S), 3 unsupervised domain adpatation (U), and 2 domain generalization (DG) methods.

Although many supervised methods have achieved high performance on large-scale datasets, like CUHK03, Market-1501 or DukeMTMC-ReID, their performance is unfortunately not satisfied on small-scale ones.
%
Many methods were proposed to deal with this problem, among which SSM~\cite{conf/cvpr/BaiBT17} and JLML~\cite{conf/ijcai/LiZG17} achieve satisfactory results. 
Nevertheless, given the limited target data, our DDAN achieves better performance since the compared methods suffer from severe over-fitting problems.

UDA methods are proposed to transfer knowledge from a large-scale labeled dataset to unlabeled ones. Some UDA approaches have shown good results for person Re-ID.
However, UDA asks for unlabeled images from datasets in the target domain and they fail to adapt to the target domain when the given training data 
are insufficient. In contrast, DDAN fully utilizes the source datasets and thus outperforms all the UDA methods in~\cref{table:result}.
%


As shown in \cref{table:result}, when we use MobileNet as the backbone network (DDAN), the Rank-1 accuracy of DDAN is $52.3\%$, $54.5\%$, $50.6\%$ and $78.5\%$ for VIPeR, PRID, GRID, and i-LIDs, respectively. It outperforms the DG method DIMN~\cite{conf/cvpr/SongYSXH19} in all the test datasets\black{, and DualNorm~\cite{journals/corr/abs-1905-03422} which has IN and BN layers in GRID and i-LIDs.} 
Note that when integrating DualNorm, a MobileNetV2 with normalization layers like~\cite{journals/corr/abs-1905-03422}, into our method and following its settings, we outperform DualNorm~\cite{journals/corr/abs-1905-03422} on all test datasets.

\begin{table}[ht]
	\centering
	\resizebox{0.5\textwidth}{!}{
		\begin{tabular}{c|c|c|c|c|c|c}
			\hline
			\multirow{2}{*}{Center domain} & \multicolumn{5}{c|}{Peripheral domains} & \multirow{2}{*}{Sum} \\ \cline{2-6}
			& Cuhk02 & Cuhk03 & Duke & Market & PersonSearch &  \\ \hline
			Cuhk02       &0     & 0.69 & 1.61 & 1.37 & 0.87 & 4.54 \\
			Cuhk03       &0.69 & 0     & 1.58 & 1.44 & 0.72 & 4.43 \\
			Duke         &1.61 & 1.58 & 0     & 1.69 & 1.20 & 6.08 \\
			Market       &1.37 & 1.44 & 1.69 & 0     & 1.10 & 5.60 \\
			PersonSearch &0.87 & 0.72 & 1.20 & 1.10 & 0     & 3.89 \\ 
			All domains & 1.81 & 1.91 & 1.92 & 1.78 & 1.93&9.35 \\\hline
		\end{tabular}
		}
	\caption{The Wasserstein distance $d_\mathrm{WS}$ between different source domains.}
	\label{table:cost}
\end{table}
\begin{table}[htb]
	\centering
	\resizebox{0.43\textwidth}{!}{
		\begin{tabular}{l|c|c|c|c}
			\hline
			Central domain & VIPeR & PRID & GRID & i-LIDs\\ \hline
            Cuhk02 & 48.4 & \textbf{48.5} & 46.6 & \textbf{74.8} \\
			Cuhk03 & 49.0 & 45.2 & \textbf{48.4} & \textbf{75.0} \\
			 Duke & 49.2 & 47.3 & 45.1 & 72.5 \\
			Market & \textbf{49.5} & 48.2 & 46.5 & 74.1\\
			PersonSearch & \textbf{50.6} & \textbf{50.0} & \textbf{47.6} & 74.6 \\  
			All domains& 48.7 & 45.4 & 44.2 & 74.5 \\\hline
		\end{tabular}
	}
	\caption{The Rank-1 accuracy (\%) of ``Baseline + $\mathcal{L}_{DA} $'' with different central domains (top-2 accuracies are \textbf{bolded}).}
	\label{table:daresult}
\end{table}
%
\subsection{Analysis} 
\label{sec:damain-wise}

\noindent
\textbf{The Effectiveness of Domain-wise Adversarial Feature Learning.}
We adopt a domain-wise adversarial loss to align the domains, which selectively reduces the gap between the peripheral domains and the central domain.
We use the baseline model to extract features from each domain, and compute the distance $d_\mathrm{WS}$ in \cref{eqn:wass} between every two domains in \cref{table:cost}. As for the pairwise alignment (All domains), since there is not a fixed central domain, we instead compute the distance between features extracted by the baseline model and the model trained with pairwise alignment. Notice that the shortest distance appears between CUHK02 and CUHK03, which is consistent with the fact that these two datasets are collected from the same location (CUHK) and thus share some sort of similarities. The PersonSearch dataset also shows a relatively small distance to the CUHK datasets, since a part of this dataset is also collected in the same location as CUHK02 and CUHK03.

\begin{figure}[tbp]
	\centering
	\subfigure[Baseline ($\mathcal{L}_\mathrm{IDE}$)]{
		\centering
		\includegraphics[width=0.49\linewidth]{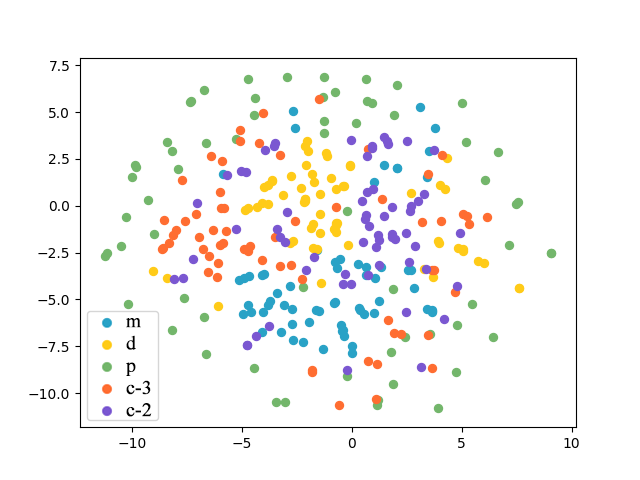}
		\label{fig:tsne_baseline}
	}%
	\subfigure[Baseline ($\mathcal{L}_\mathrm{IDE}+\mathcal{L}_\mathrm{Triplet}$)]{
		\centering
		\includegraphics[width=0.49\linewidth]{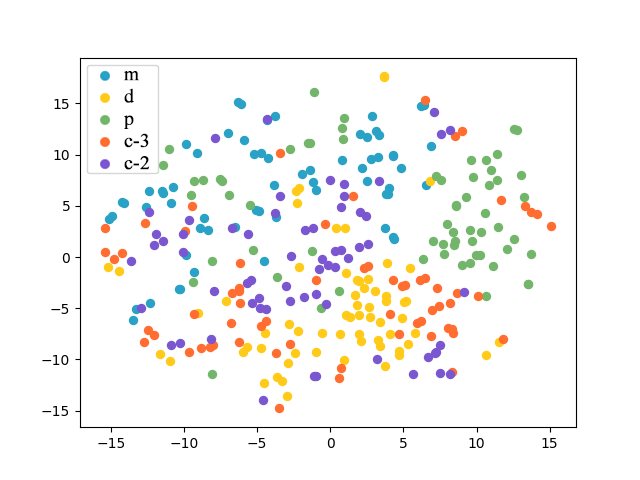}
		\label{fig:tsne_tri}
	}%
    
	\subfigure[DDAN ($\mathcal{L}_\mathrm{DA}$)]{
		\centering
		\includegraphics[width=0.49\linewidth]{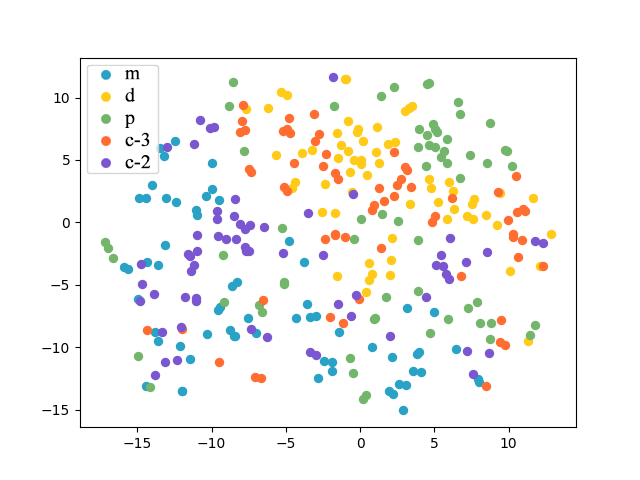}
		\label{fig:tsne_adv}
	}%
	\subfigure[DDAN ($\mathcal{L}_\mathrm{DA}+\mathcal{L}_\mathrm{SE}$)]{
		\centering
		\includegraphics[width=0.49\linewidth]{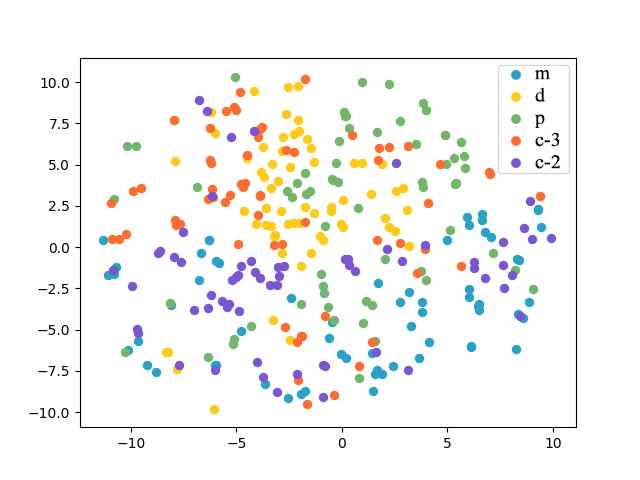}
		\label{fig:tsne_mem}
	}%
	
	\centering
	\caption{Visualization of feature distributions in all the source domains, including Market (m), Duke (d), PersonSearch (p), CUHK03 (c-3), and CUHK02 (c-2). The features are obtained by the models trained with different sets of loss functions.}
	\label{fig:tsne}
\end{figure}

In contrast, we find that a significant discrepancy appears between the Duke and Market datasets. This is consistent with our visualization results in \cref{fig:tsne_baseline}. Since the minimum cost is achieved by PersonSearch, we set it as the central domain in all the experiments.

In addition, \cref{table:daresult} shows the test performance when taking each dataset as the central domain. These results are generally consistent with the above observations in~\cref{table:cost}.
For example, when setting Duke as the central domain, the cost of aligning the peripheral domains is large. Therefore, the effectiveness of the learned the features could be hurt due to the unsuitable alignment of the peripheral domains to the central domain. Indeed, the resulting model cannot perform well on any of the four test datasets.
In contrast, the central domain PersonSearch demonstrates a better performance. In particular, the resulting rank-1 accuracy on PRID is $4.8\%$ higher than the lowest one.
Furthermore, we also evaluate the multi-domain approach (All domains) by pairwisely aligning the domains, leading to unsatisfactory performance in all the test datasets.

\begin{table}[t]
    \small
	\centering
		\begin{tabular}{c|c|c|c|c|c}
			\hline
			Image A & Image B & Image C & \multirow{2}{*}{$\mathcal{L}_{SE}$?} &\multirow{2}{*}{cs(A,B)} &\multirow{2}{*}{cs(A,C)}   \\  \cline{1-3}
			Duke & Duke & Others & & & \\ \hline
			\multirow{2}{*}{\includegraphics[width=8mm, height=12mm]{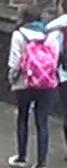}} & \multirow{2}{*}{\includegraphics[width=8mm, height=12mm]{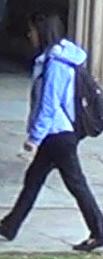}} & \multirow{2}{*}{\includegraphics[width=8mm, height=12mm]{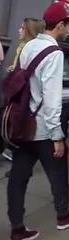}} 
			&\multirow{2}{*}{\xmark}& \multirow{2}{*}{0.67} & \multirow{2}{*}{0.59} \\[-1mm]
			&&&&&\\ \cline{4-6}
			&&&\multirow{2}{*}{\cmark}& \multirow{2}{*}{0.65} & \multirow{2}{*}{0.83} \\[-1mm]
			&&&&&\\\hline
			\multirow{3}{*}{\includegraphics[width=8mm, height=12mm]{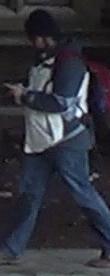}} & \multirow{3}{*}{\includegraphics[width=8mm, height=12mm]{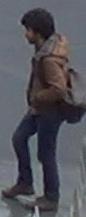}} & \multirow{3}{*}{\includegraphics[width=8mm, height=12mm]{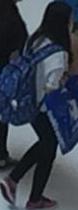}} 
			&\multirow{2}{*}{\xmark}& \multirow{2}{*}{0.78} & \multirow{2}{*}{0.75} \\[-1mm]
			&&&&&\\\cline{4-6}
			&&&\multirow{2}{*}{\cmark}& \multirow{2}{*}{0.86} & \multirow{2}{*}{0.91} \\[-1mm]
			&&&&&\\\hline
			\multirow{3}{*}{\includegraphics[width=8mm, height=12mm]{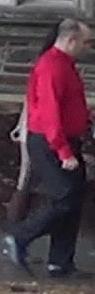}} & \multirow{3}{*}{\includegraphics[width=8mm, height=12mm]{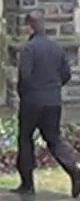}} & \multirow{3}{*}{\includegraphics[width=8mm, height=12mm]{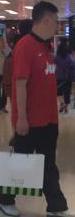}} 
			&\multirow{2}{*}{\xmark}& \multirow{2}{*}{0.76} & \multirow{2}{*}{0.71} \\[-1mm]
			&&&&&\\\cline{4-6}
			&&&\multirow{2}{*}{\cmark}& \multirow{2}{*}{0.77} & \multirow{2}{*}{0.85} \\[-1mm] 
			&&&&&\\\hline
		\end{tabular}
	\caption{The cosine similarity (cs) between representative images.}
	\label{table:identity}
\end{table}
\noindent
\textbf{The Effectiveness of Identity-wise Similarity Enhancement.}
As another contribution, the identity-wise similarity enhancement establishes the relationship between features that are visually close yet separated apart due to the domain variations. As such, it exploits multiple source domains to a greater extent.
We demonstrate the effectiveness of this component in \cref{table:identity} by comparing the cosine similarity between three representative images. In a general scenario without $\mathcal{L}_\mathrm{SE}$, this similarity is dominated by the domain variations; non-similar images (A and B) turn out to be closer than the similar ones (A and C) in the feature space, only because they are in the same domain\black{, which may have a similar hue and lighting}.
In contrast, the similarities with $\mathcal{L}_\mathrm{SE}$ can correctly reflect the relationships among A, B and C, even if the (incorrectly) large similarity between non-similar same-domain images is not penalized. \black{It successfully presents a real-world scenario, where the pedestrians in A are more likely to be the same identity as those in C than B, even if A and B are from the same domain, as long as A and C share more similar appearances than A and B.} Therefore, it captures the local similarity between A and C and effectively reduces the domain-shift.

\noindent
\textbf{Visualization.}
We use t-SNE to visualize the distribution of the features obtained by the networks with different loss functions.

%
For the baseline network with only $\mathcal{L}_\mathrm{IDE}$, the distribution in \cref{fig:tsne_baseline} shows clear discrepancy among all domains with few overlaps. Particularly, the features of Market, Duke, and PersonSearch are clearly distinguished from each other.

The triplet loss in \cref{fig:tsne_tri} shortens the intra-class distance and widens the inter-class one, thus the model learns discriminative features while also relatively increasing 
the distance between each domain, as the labels of each domain are different.
%
In particular, despite the properly aligned CUHK02 and CUHK03, the PersonSearch dataset can be seen as two parts: one well aligned with CUHK02 and CUHK03 that are collected from the same location, and the other relatively more independent one collected from movie snippets. As always, the features from Duke are distinguished from the others.

With also the domain-wise adversarial feature learning loss $\mathcal{L}_\mathrm{DA}$ in \cref{fig:tsne_adv}, the distributions of different domains are better aligned and more instances tend to be consistent with each other. However, the local distribution of Duke and part of PersonSearch are still distinguished.

Lastly, the identity-wise similarity enhancement loss in \cref{fig:tsne_mem} achieves the ideal scenario expected by DDAN, in which the distributions of similar IDs from different domains are closer. Moreover, the domain shift is greatly reduced to improve the generalization as compared against the baseline.

\begin{figure}[tbp]
	\centering
	\subfigure[Different values of $k$.]{
		\centering
		\includegraphics[width=0.48\linewidth]{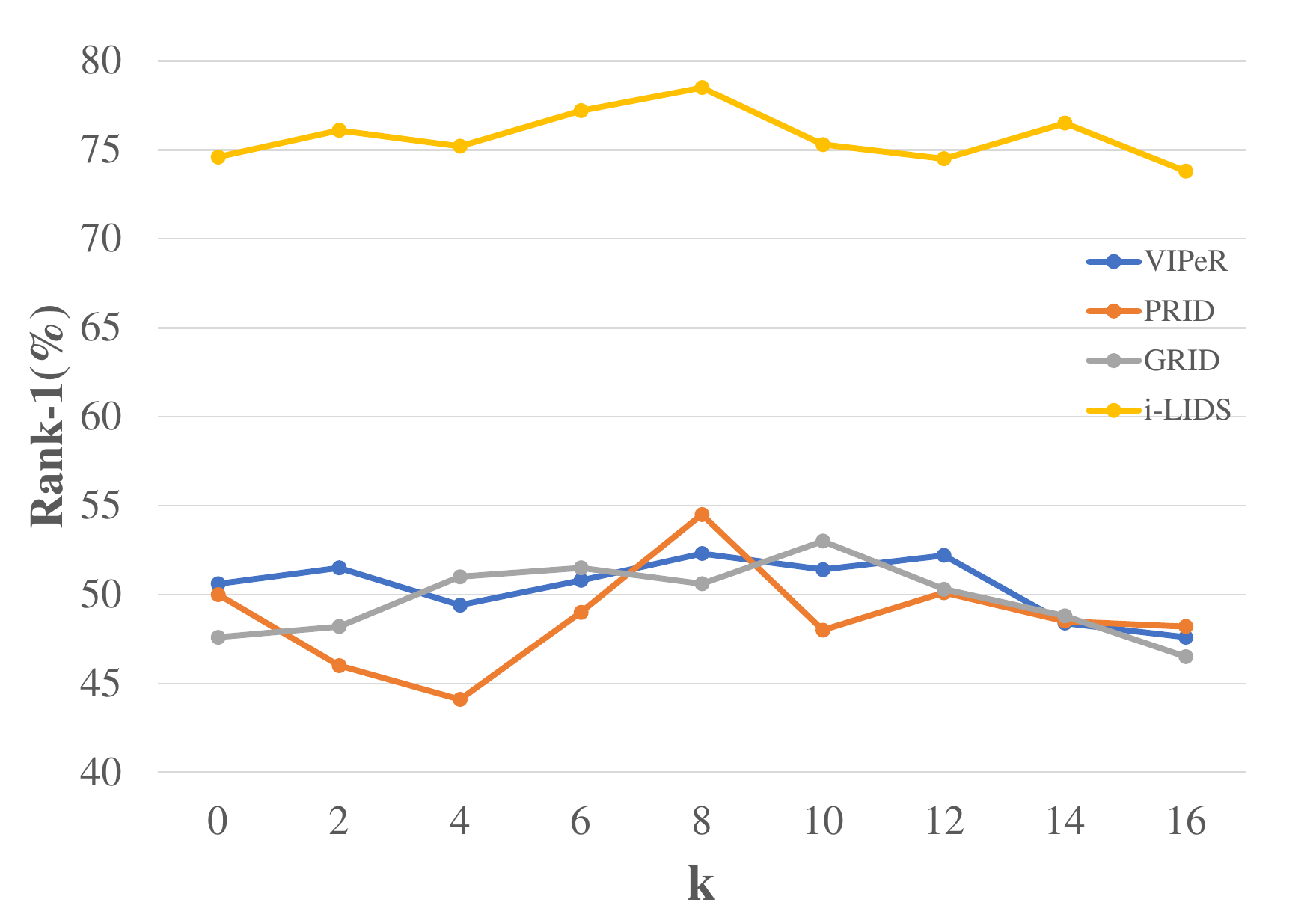}
		\label{fig:parameter_k}
	}%
	\subfigure[Different values of $\tau$.]{
		\centering
		\includegraphics[width=0.48\linewidth]{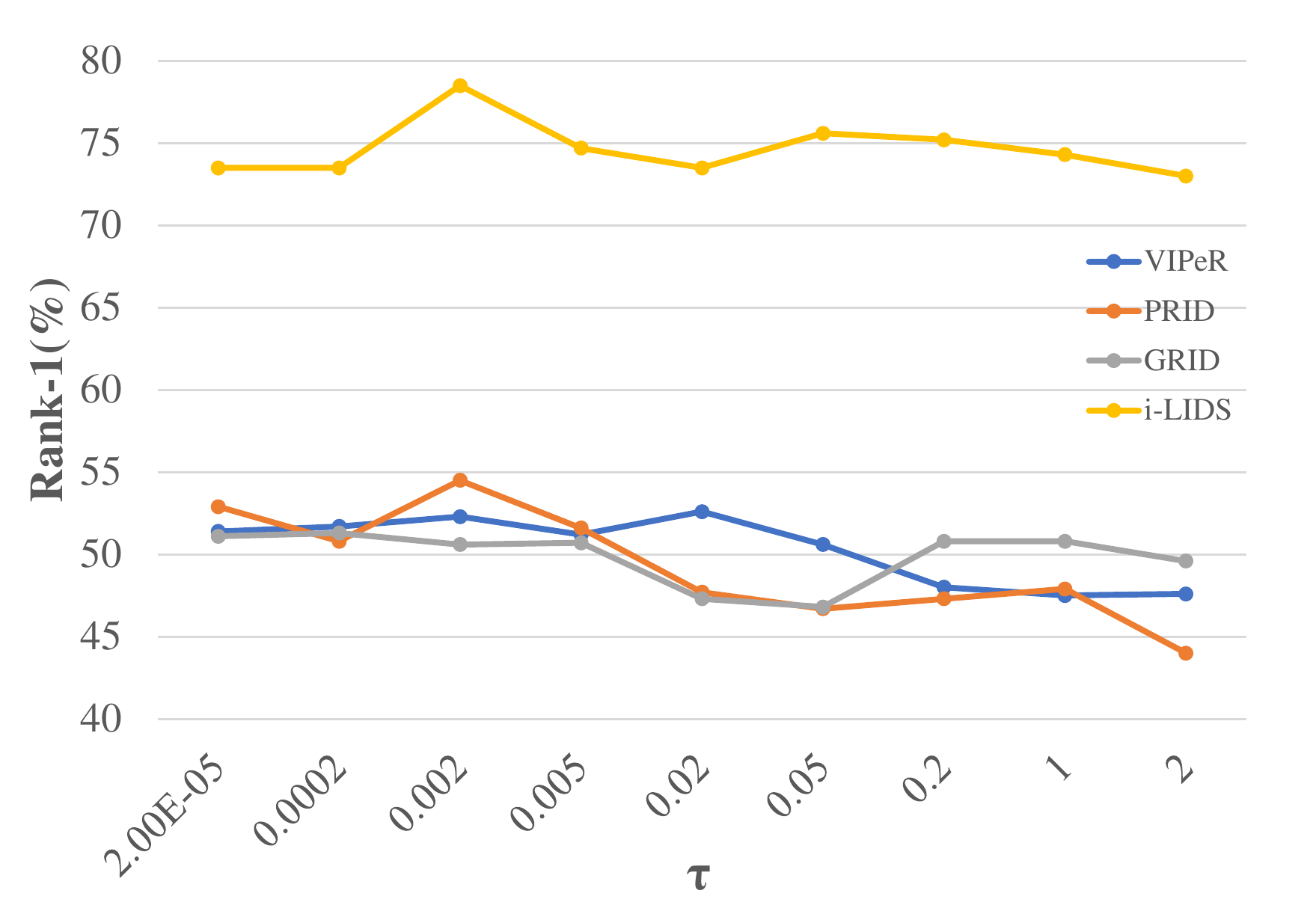}
		\label{fig:parameter_t}
	}%
	\caption{Evaluation with different values of the parameters in~\cref{eqn:kl}.}
\end{figure}
\subsection{Important Parameters}
We further explain and investigate the impact of two important hyper-parameters: the temperature $\tau$ of softmax and the number $k$ of similar IDs in~\cref{eqn:kl}. In particular, we vary the value of one parameter while fixing the other one. 

\noindent\textbf{Temperature $\tau$ of softmax.}
In \cref{eqn:kl}, we use softmax to reduce the influence of exceptionally unmatched entries appearing in the paired representations. In addition, a temperature parameter should be added to preserve the distinguishability of the  features, or the result of softmax could be small due to the large number of dimensions. We investigate the impact of the temperature $\tau$ in~\cref{fig:parameter_t} and observe that $\tau<1$ leads to a good result. However, the network cannot converge with a small $\tau$, \emph{e.g.}, $2*10^{-5}$. The best results are obtained when $\tau$ is around $2*10^{-3}$.

\noindent\textbf{Number $k$ of similar IDs.} In~\cref{fig:parameter_k}, we show the impact of using different number $k$ of similar IDs in the identity-wise similarity enhancement. For $k=0$, this enhancement is disabled. For $k\geq1$, the enabled enhancement overall improves the performance with relatively small values of $k$. 
However, a too large value of $k$ may incorrectly capture non-similar examples, which could have deleterious effects on the performance. 
Overall, $k=8$ achieves the best rank-1 accuracy and mAP in most datasets. 

\begin{table}[tbp]
	\centering
	\resizebox{\linewidth}{!}{
		\begin{tabular}{cc|c|c|c|c}
			\hline
			\multicolumn{2}{c|}{Loss functions} & VIPeR & PRID & GRID & i-LIDs\\ \hline
			$\mathcal{L}_\mathrm{IDE}$ && 41.4 & 30.8 & 38.1 & 66.2 \\
			$\mathcal{L}_\mathrm{IDE}$ + $\mathcal{L}_\mathrm{Triplet}$ && 47.2 & 46.4 & 45.3 & 72.3 \\
			$\mathcal{L}_\mathrm{IDE}$ + $\mathcal{L}_\mathrm{Triplet}$ + $\mathcal{L}_\mathrm{DA}$ && 
			50.6 & 50.0 & 47.6 & 74.6 \\
			$\mathcal{L}_\mathrm{IDE}$ + $\mathcal{L}_\mathrm{Triplet}$ + $\mathcal{L}_\mathrm{DA}$ + $\mathcal{L}_\mathrm{SE}$ && 52.3 & 54.5 & 50.6 & 78.5 \\ \hline
		\end{tabular}
	}
	\caption{Ablation study (Rank-1 accuracy (\%) is reported).}
	\label{table:ablation}
\end{table}
\subsection{Ablation Study}\label{sec:ablation}
The ablation study investigates the effectiveness of each component by adding them to the baseline one by one when being evaluated on the full test dataset.
%
As shown in~\cref{table:ablation}, $\mathcal{L}_\mathrm{Triplet}$ (with the BNNeck component~\cite{conf/cvpr/0004GLL019}) greatly improves the effectiveness of the learned representations.
$\mathcal{L}_\mathrm{DA}$ aligns the distributions of all source domains to learn a feature space that is more domain-invariant, 
resulting in the improved performance in the unseen datasets, which indicates a better generalization of learned features.
Lastly, $\mathcal{L}_\mathrm{SE}$ follows a distribution preferable in real scenario and captures identity-wise similarity to better reduce the local domain shift. 
In a word, these effective components help DDAN to learn domain-invariant features that are discriminative for the task and meanwhile insensitive to both domain- and identity-wise variations.

%

%% file: 6_conclusion.tex
\section{Conclusion}
In this paper, we proposed an end-to-end Dual Distribution Alignment Network (DDAN) to learn domain-invariant features for generalizable person Re-ID. We aligned the distributions of all available source domains at dual levels: the domain-wise adversarial feature learning and the identity-wise similarity enhancement.
%
The first one encourages to align the distributions of the peripheral domains to that of the central   domain.
Upon the determination of the central and peripheral domains, the first component largely reduces the domain discrepancy with minimum distributional shift.
The second component further reduces the local domain shift by capturing identity-wise similarity with an ID pool across the domains. 
It realizes an ideal scenario, in which any group of identities with similar visual features, though from different domains, are closer than those from the same domain. The experiments on a large-scale DG Re-ID benchmark demonstrate the superior performance of DDAN against other recent methods. 